\documentclass[12pt]{article}
\usepackage{amsmath}
\usepackage{amsthm}
\usepackage{amsfonts}
\usepackage{setspace}
\usepackage{graphicx}
\usepackage[round,comma]{natbib}
\newcommand{\disc}{\boldsymbol\eta}

\newcommand{\x}{\textbf{x}}

\newcommand{\err}{\textbf{e}}

\newcommand{\obs}{\textbf{z}}

\newcommand{\sys}{\textbf{y}}

\newcommand{\best}{\textbf{x}^{*}}
\newcommand{\ens}{\textbf{F}}
\newcommand{\ensc}{\textbf{F}_{\boldsymbol{\mu}}}
\newcommand{\bas}{\boldsymbol\Gamma}
\newcommand{\bass}{\boldsymbol\gamma}

\newcommand{\var}{\boldsymbol\Sigma}
\newcommand{\weight}{\textbf{W}}
\newcommand{\B}{\textbf{B}}

\newcommand{\cc}{\textbf{c}}

\newcommand{\R}{\mathcal{R}}

\newcommand{\ibound}{T}

\newcommand{\fimpl}{\mathcal{I}}
\newcommand{\cimpl}{\tilde{\mathcal{I}}}
\providecommand{\keywords}[1]{\textbf{\textit{Keywords:}} #1}
\newtheorem{theorem}{Theorem}
\newcommand{\mat}{\boldsymbol{\Psi}}
\newcommand{\bmu}{\boldsymbol{\mu}}

\begin{document}

\title{Efficient calibration for high-dimensional computer model output using basis methods}
\author{James M. Salter$^1$\thanks{
			\textit{The authors gratefully acknowledge support from EPSRC fellowship No. EP/K019112/1, and would also like to thank the Isaac Newton Institute for Mathematical Sciences, Cambridge, for support and hospitality during the Uncertainty Quantification programme where work on this paper was undertaken (EPSRC grant no EP/K032208/1).}}  \& Daniel B. Williamson$^{1,2}$ \\ $^1$Department of Mathematics, University of Exeter, Exeter, UK. \\$^2$Alan Turing Institute, London, UK.}
\maketitle


\begin{abstract}
Calibration of expensive computer models with high-dimensional output fields can be approached via history matching. If the entire output field is matched, with patterns or correlations between locations or time points represented, calculating the distance metric between observational data and model output for a single input setting requires a time intensive inversion of a high-dimensional matrix. By using a low-dimensional basis representation rather than emulating each output individually, we define a metric in the reduced space that allows the implausibility for the field to be calculated efficiently, with only small matrix inversions required, using projection that is consistent with the variance specifications in the implausibility. We show that projection using the $L_2$ norm can result in different conclusions, with the ordering of points not maintained on the basis, with implications for both history matching and probabilistic methods. We demonstrate the scalability of our method through history matching of the Canadian atmosphere model, CanAM4, comparing basis methods to emulation of each output individually, showing that the basis approach can be more accurate, whilst also being more efficient.
\end{abstract}


\section{Introduction}

A computer model, $f(\cdot)$, is a representation of a real-world process, given by a set of equations and parametrisations, that takes a vector of inputs $\x$, and returns an output. This output may be a single value, a spatial field, a time series, or a combination of these across multiple different fields (e.g. climate models \citep{von2013canadian}). Computer models often represent complex processes, and may require long running times on expensive supercomputers. It is therefore only possible to evaluate the model at a small sample of values from the input space.
\par
Statistical models (`emulators') are commonly used as a proxy for expensive computer models, giving predictions for the output at unseen values of $\x$, along with an uncertainty where the true model has not been run \citep{sacks1989design, higdon2008computer}. Such emulators can then be used to calibrate the inputs, $\x$, of the computer model, based on observations of the real-world process. This can be done either probabilistically, with a distribution given for the best setting of the input parameters (`Bayesian calibration', \citet{kennedy2001bayesian}), or via history matching \citep{craig1996bayes, williamson2015bias, andrianakis2017efficient}.
\par
History matching, unlike probabilistic calibration, does not require any distributional assumptions and, instead of returning a distribution, rules out regions of the input parameter space that are inconsistent with the observations, based on an implausibility measure. Performed iteratively (`refocussing', \citet{vernon2010galaxy}, \citet{williamson2017tuning}), history matching is a powerful tool, allowing the region of space that leads to output consistent with the observations to be identified, if it exists (unlike probabilistic calibration, the result of history matching can be that there are no settings of $\x$ that give model output consistent with the observations, up to observation error and model discrepancy). Refocussing is performed by selecting a new design of points from the current not implausible region of space, and running these on the expensive model, before repeating the emulation and history matching process given these new evaluations of the computer model. By performing more iterations (`waves'), the density of points in the reduced, not ruled out, space increases, and if probabilistic calibration were then performed in this space, the results are usually more accurate than if it were immediately implemented over the full input space \citep{vernon2010galaxy, salter2016comparison, salter2018uncertainty}.
\par
High-dimensional computer model output has several different forms, requiring different approaches in order to emulate the output. For example, time series output often lends itself to an autoregressive approach \citep{liu2009dynamic, williamson2014evolving}, whilst spatial fields are often projected onto a low-dimensional basis given by the principal components of the output \citep{higdon2008computer, chang2016calibrating}, or some other optimally-selected basis for calibration \citep{salter2018uncertainty}. In these cases, emulators are then fitted for the coefficients in the reduced space. This reduced basis approach may be used for temporal or spatio-temporal output with few or no adjustments required (\citet{higdon2008computer} demonstrate the method using a spatio-temporal example). The low-dimensional basis method is attractive because it reduces the dimensionality, and hence computation time required, when the field dimension is very large, whilst also maintaining interpretability, and correlations, from the full output through the basis vectors.
\par
An alternative approach towards emulating spatial and spatio-temporal output is to emulate every grid box or time point individually \citep{lee2012mapping, spiller2014automating, gu2016parallel, johnson2018importance}. As the number of emulators to be built scales with the size of the output, \citet{gu2016parallel} set common regressors for the mean function, and fix the correlation parameters, across all grid boxes. Validating emulators for thousands of grid boxes may be a challenge, and only an automated approach to this is generally feasible.
\par
Given a set of emulators for the model output, how best to overcome the problem of high-dimensionality when calibrating these fields is not clear. One approach is to use emulated coefficients to reconstruct the original field, and compare it to the observations themselves \citep{wilkinson2010bayesian}, although as the dimension of the field increases, this becomes increasingly intractable. Instead, all quantities defined over the field can be projected onto a low-dimensional basis, with the representation of the observations compared to emulated output on the basis \citep{higdon2008computer, sexton2011multivariate, chang2016calibrating}, with fast calculations in this reduced subspace.
\par
In this paper, we show that whilst history matching using a low-dimensional representation of the original output is often necessary for computational reasons, the standard method for setting the bound to define the space of not ruled out points may lead to results that are not consistent with those over the field. We provide an efficient way to calculate the implausibility over the original field, so that history matching high-dimensional fields is tractable, only requiring evaluations of the inexpensive coefficient implausibility. This simplification requires projection in a certain norm, and we demonstrate the importance of projection in the `correct' norm, with the ordering of points according to the distance metric varying between the full space and subspace, and in fact the same parameter choice not minimising the two measures, with different projection choices. Due to the relationship between the implausibility and the likelihood, this result also has implications for probabilistic calibration if directly performed in a subspace. We also compare basis emulation methods to univariately emulating each grid box, demonstrating the computational savings afforded by using a basis, without sacrificing performance, with the basis method outperforming the univariate method in our climate application.
\par
Section \ref{litreview} outlines emulation and history matching for high-dimensional fields. Section \ref{sectionprojection} describes history matching in the projected space, with this demonstrated via a high-dimensional example, the CanAM4 climate model, in Section \ref{sectexample}. Section \ref{sectionefficient} shows how the field implausibility can be calculated dependent on small matrix inversions, given a certain choice of projection. Section \ref{sectionapplication} fits basis and univariate emulators to the sea level pressure field of CanAM4, with the `emulate every grid box' approach to calibration compared with the fast reduced basis method. Section \ref{discussion} contains discussion.

\section{Spatio-temporal history matching} \label{litreview}

History matching rules out settings of the input parameters, $\x \in \mathcal{X}$, that lead to computer model output, $f(\x)$ (a vector of length $\ell$), that are not consistent with observations, $\obs$, given an error specification \citep{craig1996bayes, williamson2013history, andrianakis2017efficient, vernon2018bayesian}. History matching uses a statistical model that links the true value of the system, $\sys$, with the computer model, generally given by \citep{kennedy2001bayesian}:
\begin{equation} \label{hmeqn}
\obs = f(\best) \oplus \disc \oplus \err,
\end{equation}
where $\disc$ (the discrepancy between the output given at the `best' setting, $\best$, of $f(\cdot)$, and reality, $\sys$) and $\err$ (the observation error) are uncorrelated (indicated by $\oplus$) mean-zero terms, with positive definite variance matrices $\var_{\disc}$ and $\var_{\err}$ respectively. Rather than requiring full distributions on $\disc$ and $\err$, as for probabilistic calibration, history matching only uses expectations and variances. When $f(\cdot)$ is expensive to run, it is replaced by an emulator (Section \ref{sectionem}).
\par
The implausibility, $\mathcal{I}(\x)$, for a parameter setting $\x$ is defined as the Mahalanobis distance between the observations and the predictive expectation from an emulator for the computer model:
\begin{equation} \label{implmv}
\mathcal{I}(\x) = (\obs - \text{E}[f(\x)])^{T}(\text{Var}(\obs - \text{E}[f(\x)]))^{-1} (\obs - \text{E}[f(\x)]),
\end{equation}
where, under the model assumptions in \eqref{hmeqn}, we have:
\begin{equation} \label{hmvar}
\text{Var}(\obs - \text{E}[f(\x)]) = \text{Var}[f(\x)] + \var_{\err} + \var_{\disc}.
\end{equation}
Large values of this distance indicate that it is implausible that $\x = \best$. Using $\mathcal{I}(\x)$, `Not Ruled Out Yet' (NROY) space contains all not implausible $\x$, defined as \citep{vernon2009bayes,vernon2010galaxy}:
\begin{displaymath}
\mathcal{X}_{NROY} = \{\x \in \mathcal{X} | \mathcal{I}(\x) < \ibound \},
\end{displaymath}
for bound $\ibound$. If $\mathcal{I}(\x) \sim \chi^2_{\ell}$, for $\ell$ the rank of \eqref{hmvar}, then $\ibound = \chi^2_{\ell, 0.995}$, so that $P(\mathcal{I}(\x) < \ibound) = 0.995$.

\subsection{Emulation} \label{sectionem}

Emulators are used in place of the computer model when it is costly or time-consuming to run, with Gaussian processes a popular choice \citep{sacks1989design,haylock1996inference,salter2016comparison}. 
Emulation depends on having run the true model, $f(\cdot)$, at $n$ settings $\x \in \mathcal{X}$, giving ensemble $\ens = (f(\x_1), \ldots, f(\x_n))$ , with $f(\x_i)$ an $\ell$-dimensional vector.

\subsubsection{Univariate emulators}

In this setting, each of the $\ell$ outputs of $f(\cdot)$, denoted by subscript $i$, is emulated as a Gaussian process, with:
\begin{displaymath}
f_i(\x) \sim \text{GP}(m_i(\x), R_i(\x, \x)), \quad i = 1, \ldots, \ell,
\end{displaymath}
for mean function $m_i(\cdot)$, and covariance $R_i(\cdot, \cdot)$. These functions may be fitted individually for all $\ell$ outputs, allowing different terms in the mean function, and different correlation lengths \citep{lee2013magnitude, spiller2014automating, johnson2018importance}, or, for computational convenience, a fixed set of regressors may be imposed across all $\ell$ outputs, with a single set of correlation lengths estimated \citep{gu2016parallel}. The former approach offers greater flexibility, although is more time consuming.

\subsubsection{Basis emulation}

For validation and computational purposes, low-dimensional representations of the output are commonly used, with significantly fewer emulators than a univariate approach required. The high-dimensional data is projected onto a basis, often given by the principal components across the model runs (the Singular Value Decomposition (SVD) basis) \citep{higdon2008computer, sexton2011multivariate, chang2014probabilistic}. 
\par
To find the SVD basis, the ensemble mean, $\bmu$, given by averaging across the rows of $\ens$, is subtracted from each column of $\ens$, to give the centred ensemble, $\ensc$. The SVD basis, $\bas$, is defined as:
\begin{equation} \label{svdeqn}
\ensc^T = \textbf{U} \textbf{D} \bas^T.
\end{equation}
The basis is truncated after the first $q$ vectors, for truncated basis $\bas_q = (\bass_1, \ldots, \bass_q)$ sufficient to explain a high (commonly, 90\% or 95\%, but problem dependent) proportion of the variability in $\ensc$. Projection of an output field, $f(\x)$, onto basis $\bas_q$ is given by:
\begin{displaymath}
\textbf{c}(\x_{i}) = (\bas_q^{T}\weight^{-1}\bas_q)^{-1} \bas_q^{T} \weight^{-1} f(\x_{i}),
\end{displaymath}
for a positive definite weight matrix $\weight$. We discuss the role of $\weight$ later. If $\weight \propto \mathbb{I}_{\ell}$, this is the SVD ($L_2$) projection. A set of coefficients is mapped back to the dimension of the original field using:
\begin{displaymath}
f(\x_{i}) = \bas_q \textbf{c}(\x_{i}) + \boldsymbol{\epsilon},
\end{displaymath}
for error vector $\boldsymbol{\epsilon}$. If $q = n$, then $\boldsymbol{\epsilon} = \textbf{0}$ for $\x_i \in \textbf{X} = (\x_1, \ldots, \x_n)$.
\par
Emulators are built for the coefficients on the first $q$ basis vectors,
\begin{displaymath}
c_i(\x) \sim \text{GP}(m_i(\x), R_i(\x, \x)), \quad i = 1, \ldots, q,
\end{displaymath}
with $\text{E}[\cc(\x)] = (\text{E}[c_1(\x)], \ldots, \text{E}[c_q(\x)])^T$, the emulator expectation for each of the $q$ basis vectors, and $\text{Var}[\cc(\x)] = diag(\text{Var}[c_1(\x)], \ldots, \text{Var}[c_q(\x)])$ the associated $q \times q$ variance matrix. Alternatively, a multivariate emulator could be fitted for the $q$ coefficients, with $\text{Var}[\cc(\x)]$ potentially containing covariances between the coefficients (orthogonality of basis vectors does not imply independence of projected coefficients). In either case, we retrieve the $\ell$-dimensional expectation and variance of $f(\x)$ via:
\begin{equation} \label{fieldem}
\text{E}[f(\x)] = \bas_q \text{E}[\cc(\x)], \quad \text{Var}[f(\x)] = \bas_q \text{Var}[\cc(\x)] \bas_q^T.
\end{equation}

\subsubsection{Basis rotation}

The space of possible reconstructions, $\bas_q \cc(\x)$, is a $q$-dimensional surface in $\ell$-dimensional space, restricted by the basis. \citet{salter2018uncertainty} show that using the SVD basis, without considering the observations (i.e. what we want the model to be able to reproduce, if possible) can lead to guaranteeing that the conclusion of a calibration exercise is that the computer model cannot represent $\obs$, regardless of whether this is true (the `terminal case'). To avoid this, prior to building emulators for a given basis, we consider the `reconstruction error', a measure of how accurately the observations can be represented by a basis $\B_q$ \citep{salter2018uncertainty}:
\begin{equation} \label{reconerror}
\R_{\weight}(\B_q, \obs) = \lVert \obs - \B_q (\B_q^{T} \weight^{-1} \B_q)^{-1} \B_q^{T} \weight^{-1} \obs \rVert_{\weight},
\end{equation}
for $\ell \times \ell$ positive definite weight matrix $\weight$, and where $\lVert \textbf{v} \rVert_{\weight} = \textbf{v}^{T} \weight^{-1} \textbf{v}$ is the norm of vector $\textbf{v}$.  By setting $\weight = \var_{\err} + \var_{\disc}$, $\R_{\weight}(\B_q, \obs)$ is equivalent to $\fimpl$ if the emulator variance $ \text{Var}[f(\x)]$ = \textbf{0}. Therefore, if $\R_{\weight}(\B_q, \obs) > T$, then the representation of $\obs$ on the basis would be ruled out. If this is true, we search for an optimal rotation that reduces the reconstruction error.
\par
The rotation is found by iteratively selecting linear combinations of the SVD basis, $\bas$, combining important patterns for explaining the observations with patterns that explain ensemble variability, so that emulators can be built. By ensuring that the observations are explained as well as allowed by the ensemble $\ens$, we are able to potentially identify input parameters that lead to the computer model reproducing $\obs$, and avoid guaranteeing that we will rule these out. Full details are given in \citet{salter2018uncertainty}.

\section{History matching with large $\ell$} \label{sectionprojection}

We want to calibrate using all available information, incorporating any knowledge about correlations from $\var_{\err}$, $\var_{\disc}$, and the model output into the resulting analysis. As $\ell$ increases, calculating $\fimpl(\x)$ (equation \eqref{implmv}), which does include all information about the $\ell$-dimensional field, becomes exponentially more expensive, due to the necessary inversion of an $\ell \times \ell$ variance matrix that varies with $\x$. To history match, the implausibility must be evaluated thousands or millions of times, particularly if either several waves are performed, or if the resulting NROY space is small, so that it is difficult to sample from \citep{williamson2013efficient,andrianakis2015bayesian}. Similarly, probabilistic calibration requires repeated evaluations of the likelihood within an MCMC sampler, resulting in the same computational problem.
\par
For large $\ell$, therefore, calculating $\fimpl(\x)$ is not currently feasible, and instead, it is attractive to apply a low-dimensional basis approach to emulation and calibration. Given a basis, $\bas_q$, and emulators for the coefficients on these $q$ basis vectors, we can history match in the subspace defined by $\bas_q$, as has been performed extensively for probabilistic calibration \citep{higdon2008computer, sexton2011multivariate, chang2014probabilistic, chang2016calibrating}. We define the `coefficient implausibility', analogous to \eqref{implmv} in the subspace, as:
\begin{equation} \label{coeffimpl}
\cimpl_{\weight}(\x) = (\cc(\obs) - \text{E}[\cc(\x)])^{T}(\text{Var}[\cc(\x)] + \text{Var}[\cc(\err)] + \text{Var}[\cc(\disc)])^{-1} (\cc(\obs) - \text{E}[\cc(\x)]),
\end{equation}
where subscript $\weight$ indicates that projection of $\ell$-dimensional quantities is performed in matrix norm $\weight$, for positive definite $\weight$, i.e. $\obs$, $\var_{\err}$ and $\var_{\disc}$ are projected onto basis $\bas_q$ as follows (see \citet{salter2018uncertainty} for proof that this projection is optimal):
\begin{align*}
\begin{split}
\textbf{c}(\obs) &= (\bas_q^{T}\weight^{-1}\bas_q)^{-1} \bas_q^{T} \weight^{-1} \obs, \\
\text{Var}[\textbf{c}(\err)] &= (\bas_q^{T}\weight^{-1}\bas_q)^{-1} \bas_q^{T} \weight^{-1} \var_{\err} \weight^{-1} \bas_q (\bas_q^{T}\weight^{-1}\bas_q)^{-T}, \\
\text{Var}[\textbf{c}(\disc)] &= (\bas_q^{T}\weight^{-1}\bas_q)^{-1} \bas_q^{T} \weight^{-1} \var_{\disc} \weight^{-1} \bas_q (\bas_q^{T}\weight^{-1}\bas_q)^{-T}. \\
\end{split}
\end{align*}
The measure in \eqref{coeffimpl} requires only $q \times q$ matrix inversions, with $q << \ell$, hence history matching a large spatial field becomes tractable.
\par
Whether we project with $\weight = \mathbb{I}_{\ell}$ ($L_2$ projection), as is often the case for applications that use the SVD basis, or $\weight = \var_{\err} + \var_{\disc}$, as in \citet{higdon2008computer} and \citet{salter2018uncertainty} (in the latter, for consistency with the reconstruction error \eqref{reconerror} and rotation), will affect the resulting NROY space. In the $L_2$ case, we are treating all regions of the output field equally when projecting, whereas with an alternative $\weight$, weightings of space given by the discrepancy and observation error variances are reflected when $\obs$, $\var_{\err}$, $\var_{\disc}$ and $\ensc$ are projected onto $\bas_q$. We demonstrate the impact that the choice of projection can have in Section \ref{sectexample}.

\subsection{Canadian climate model} \label{canam4}

CanAM4 is an atmosphere-only global climate model \citep{von2013canadian} with many output fields that could be used in a calibration exercise. When history matching, we can choose to initially use a subset of the output fields, and rule out regions of parameter space that are inconsistent for this subset alone (whereas in probabilistic calibration, we would need to emulate and calibrate all outputs of interest simultaneously). Here, we consider the sea level pressure (SLP) field, given on a $128\times64$ longitude-latitude grid ($\ell = 8192$). The ensemble, $\ens = (f(\x_1), \ldots, f(\x_n))$, has $n = 62$ members, obtained by running CanAM4 at a space-filling design in the 13-dimensional input space, $\mathcal{X}$ \citep{williamson2015exploratory}.
\par
We are not able to run further ensembles of CanAM4, due to the supercomputer time required for running a GCM. Throughout this article, to assess the accuracy of emulation and calibration methods, we use proxy observations, given by a run from a second 49 member ensemble of CanAM4 (from the application in \citet{salter2018uncertainty}), so that we know there are input parameters, $\best$, such that the `observations' can be produced by the climate model. This run is plotted in the top left panel of Figure \ref{pslobs}, with the ensemble mean, $\bmu$, subtracted, showing how this run is generally different from $\ens$. There are positive biases over Asia, North America, and Antarctica, with negative biases in the North Atlantic and Pacific.

\subsection{Example} \label{sectexample}

Using CanAM4, we now demonstrate the difference that can be caused by the projection choice. In this section, rather than building emulators, we use the ensemble output itself, demonstrating that any problems are independent of emulator quality (hence $\text{Var}[\cc(\x)] = 0$ in all implausibility calculations). As we know that $\best$ exists, the discrepancy variance, $\var_{\disc}$, is equal to 0. We use $\var_{\err}$ to represent our tolerance to error (how close runs should be to the proxy observations to be deemed acceptable). We define $\var_{\err}$ as a Gaussian covariance matrix (dependent on longitude and latitude), with $(i,j)^{th}$ entry:
\begin{equation} \label{obserr}
\var_{\err}^{ij} = \sigma_i \sigma_j \mathrm{exp} \{ -(\frac{lon_i - lon_j}{\delta_{lon}})^{2} - (\frac{lat_i - lat_j}{\delta_{lat}})^{2} \},
\end{equation}
and vary the correlation lengths, $\boldsymbol{\delta} = (\delta_{lon}, \delta_{lat})$, to alter the correlation between close locations, and the standard deviations, $\sigma_i$, to give different weightings of the output space. Increasing $\boldsymbol{\delta}$, and having non-constant $\sigma_i$, has the effect of making $\var_{\err}$ less similar to the identity matrix.
\par
We calculate the SVD basis across the ensemble (equation \eqref{svdeqn}), and truncate after 90\% of variability is explained to give basis $\bas_q$. Using the coefficient implausibility (equation \eqref{coeffimpl}), we compare $\cimpl_{L_2}$ (projection with $\weight = \mathbb{I}_{\ell}$) and $\cimpl_{\weight}$, for several different choices of $\weight = \var_{\err}$. The $L_2$ projection treats all outputs equally, but $\cimpl_{L_2}$ does involve $\var_{\err}$, incorporating any structure given by this variance matrix, so that it is not immediately clear that the resulting implausibilities will be substantially different, as the only difference is given by the projection method. Using $\bas_q$, and a choice of $\var_{\err}$, we calculate $\cimpl_{L_2}$ and $\cimpl_{\weight}$ for the 62 ensemble members, and compare the two measures.
\par
\begin{figure}[t]
\centering
\includegraphics[width = 0.8\linewidth]{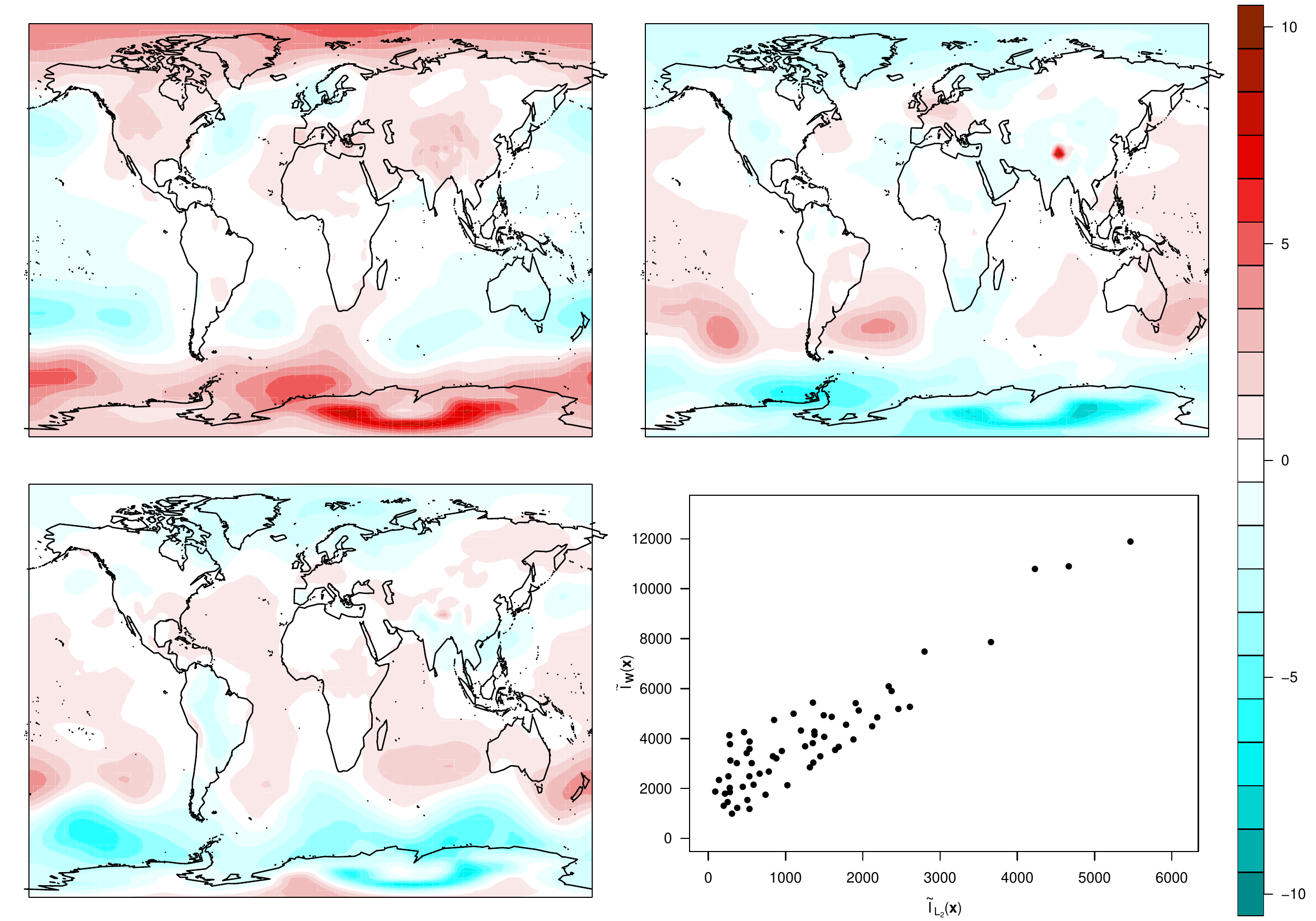}
\caption{Top left: the proxy observations for sea level pressure, relative to the ensemble mean. Top right/bottom left: the ensemble members minimising $\cimpl_{\weight}$ and $\cimpl_{L_2}$ respectively, relative to $\obs$. Bottom right: $(\cimpl_{L_2}, \cimpl_{\weight})$ for every ensemble member, with $\boldsymbol{\delta} = (5,5)$ and higher error tolerance in Antarctica.}
\label{pslobs}
\end{figure}
First, we set $\boldsymbol{\delta} = (5,5)$, giving some correlated structure to $\var_{\err}$, and set $\sigma_i = 10$ for latitude below 54$^{\circ}$S, and $\sigma_i = 1$ elsewhere, to allow greater tolerance to errors around Antarctica. Figure \ref{pslobs} shows the runs that minimise each implausibility measure across the ensemble, and compares the two measures for the 62 model runs. The best runs are not the same, with the ensemble member that minimises $\cimpl_{\weight}$ (top right) generally lower than $\obs$ over land, and higher over the oceans, whilst the minimiser for $\cimpl_{L_2}$ (bottom left) is mostly higher than $\obs$, with the exception of high latitudes. There is not consistent ordering between each measure, leading to potential differences in the composition of NROY space, and in the posterior distribution for $\best$ in a calibration exercise. Although there is a positive relationship between the measures, there is a wide range of $\cimpl_{L_2}$ values associated with a given $\cimpl_{\weight}$.  Calculating the implausibility on the field (equation \eqref{implmv}) suggests that, given this specification of $\var_{\err}$, the run that minimises $\cimpl_{\weight}$ is superior.
\par
To further demonstrate the problem, we now select two subsets of the spatial output, each with $\ell = 800$: region 1, with longitude from 14-124$^{\circ}$W and latitude between 21$^{\circ}$S and 32$^{\circ}$N, and region 2 (same longitude, latitude above 35$^{\circ}$N), each covering some part of the Americas. For each region, we calculate the truncated SVD basis, with $q = 10$ and $q = 8$ respectively.
\par
\begin{figure}[t]
\centering
\includegraphics[width = 0.7\linewidth]{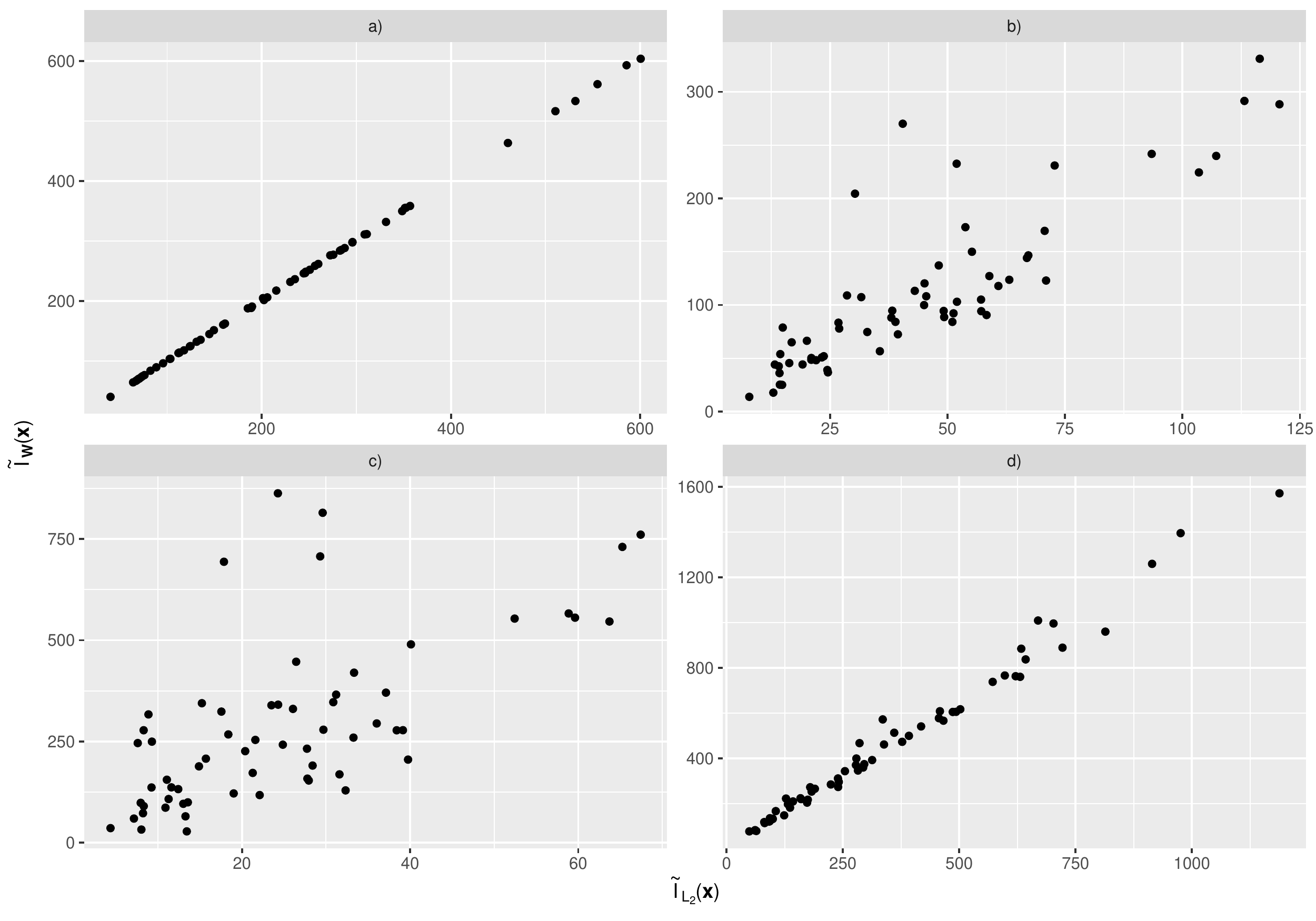}
\caption{Plots of $\cimpl_{\weight}(\x)$ against $\cimpl_{L_2}(\x)$ for the CanAM4 ensemble: a) region 1, $\boldsymbol{\delta} = (2,2)$; b) region 1, $\boldsymbol{\delta} = (5,5)$; c) region 1, $\boldsymbol{\delta} = (10,5)$; d) region 2, $\boldsymbol{\delta} = (10,5)$.}
\label{coeffexample}
\end{figure}

Each example proceeds as before, with $\cimpl_{L_2}$ and $\cimpl_{\weight}$ calculated for each ensemble member, for some choice of $\weight = \var_{\err}$. Figure \ref{coeffexample} compares $\cimpl_{L_2}$ and $\cimpl_{\weight}$ for various choices of $\boldsymbol{\delta}$ and region (with $\sigma_i = 1 \, \forall i$). The first three panels relate to region 1, showing the effect that increasing the correlation in $\var_{\err}$ can have. In the first panel, $\boldsymbol{\delta} = (2,2)$, and the two measures are almost perfectly correlated. Increasing the correlation lengths to $\boldsymbol{\delta} = (5,5)$ (panel b)), there is no longer a near-perfect relationship between the two measures. In this example, the same ensemble member minimises each implausibility, however the general ordering of runs is different.
\par
Panel c) shows the implausibilities if the correlation in longitude only is now increased ($\boldsymbol{\delta} = (10,5)$). This reduces the correlation between the two measures further, and now the runs that minimise each are different (the run that minimises $\cimpl_{\weight}$ has only the 19th lowest $\cimpl_{L_2}$ value). Panel d) uses the same $\var_{\err}$ as c), but for region 2: the relationship between the two measures is not only dependent on $\weight$, but also on the ensemble, and hence the basis.
\par

These examples collectively highlight several problems. First, the ordering of points in each distance measure is not the same, hence the `best' setting of the inputs, $\best$, can be different, dependent on the projection method. It follows from this that NROY space is likely to have a different composition, with increasing correlation in $\weight$ generally resulting in a greater difference. The lack of consistent ordering also has implications for probabilistic calibration: the calibration likelihood contains a similar calculation as the implausibility, so that the posterior distribution for $\best$ (and potentially the value of $\x$ that maximises the likelihood) will be affected by changing the relative likelihoods of points. There is no general relationship between the level of correlation in $\weight$, and the consistency between the measures (example d)). 
\par
The examples also potentially show the inadequacy of the chi-squared bound, with substantially different ranges of implausibilities across the ensemble, dependent on whether projection was in $L_2$ or $\weight$. For c), $\cimpl_{L_2}(\x)$ has a maximum across the ensemble of 68, whereas $\cimpl_{\weight}(\x)$ has a maximum of over 800. The bound in this case is $\chi^2_{q, 0.995} = 25.2$, either ruling out around half of the ensemble ($\cimpl_{L_2}(\x)$) or the whole ensemble ($\cimpl_{\weight}(\x)$). Calculating the field implausibility instead for $\var_{\err}$ as in c), the majority of runs are not ruled out with $\chi^2_{\ell, 0.995}$, inconsistent with both coefficient implausibilities. Therefore, the chi-squared bound for the subspace is not analogous to that on the field, i.e., the normality assumption does not hold in the projection space, despite the equivalent form of the measure used.
\par
In conclusion, the choice of projection can lead to different results, independently of the emulator type or quality, if we calibrate a high-dimensional field using a subspace projection.

\section{Efficiently calculating $\fimpl(\x)$} \label{sectionefficient}

Exploiting a basis structure, we can efficiently calculate the original implausibility, $\fimpl$, demonstrating which basis projection method should be used. We show that $\fimpl$ can be decomposed so that only a single expensive inversion of an $\ell \times \ell$ matrix is required, with all of the variability due to $\x$ evaluated within a $q \times q$ inversion, given an appropriate choice of $\cimpl_{\weight}$. The proof (given in the Appendix) relies on the well-known Woodbury formula \citep{woodbury1950inverting, higham2002accuracy}, which is also used for efficient calculations by \citet{higdon2008computer} (for inverting the high-dimensional matrix in the calibration likelihood) and \citet{rougier2008efficient} (outer product emulation).
\begin{theorem}
For basis $\bas_q$, and $\weight = \var_{\err} + \var_{\disc}$, we have:
\begin{equation} \label{implequation}
\fimpl(\x) = \R_{\weight}(\bas_q, \obs) + \cimpl_{\weight}(\x).
\end{equation}
\end{theorem}
\noindent That is, we have the reconstruction error given by the truncated basis, and the coefficient implausibility with $\weight = \var_{\err} + \var_{\disc}$. As shown by the examples in Section \ref{sectexample}, how to project into the $q$-dimensional subspace affects the results. Here, we see that projection with this $\weight$ is the appropriate choice, as this gives consistency with $\fimpl$, the implausibility over the field.
\par
By setting $\weight = \var_{\err} + \var_{\disc}$, we have consistency between the method of projection, and the implausibility metric. Projecting using this $\weight$, the field implausibility at $\x$ can be written as the sum of $\R_{\weight}(\bas_q, \obs)$, the reconstruction error \eqref{reconerror} of $\obs$ on basis $\bas_q$ (fixed for all $\x \in \mathcal{X}$), and $\cimpl_{\weight}(\x)$, the $\weight$-projected implausibility on the basis at $\x$, involving only $q$-dimensional matrix multiplications, for small $q$. Hence, by projecting each $\ell$-dimensional quantity using $\weight = \var_{\err} + \var_{\disc}$, we can find $\fimpl(\x)$ for any $\x$, given a one-off expense of inverting $\weight$ for the reconstruction error. In order to history match, $\var_{\err}$ and $\var_{\disc}$ must be set regardless, so that also requiring these for projection is not restrictive. 
\par
It is not important that the basis be orthogonal in $L_2$, $\weight$, or with respect to any other norm. The emulators for the coefficients are dependent on the basis choice, but as this remains fixed, orthogonality is not key, and coefficients projected onto orthogonal vectors are not necessarily uncorrelated. We are aiming to predict the output at $\x$ over the original field, and the emulation of coefficients is a method for obtaining this field prediction, with Theorem 1 allowing emulated coefficients to be compared to $\obs$.
\par
As we can now calculate implausibility for the $\ell$-dimensional field, the observation that the chi-squared bound may not be suitable on the basis, as seen in Section \ref{sectexample}, is not an issue. We note that if the observations can be represented perfectly by the basis ($\R_{\weight}(\bas_q, \obs) = 0$), then $\fimpl(\x) = \cimpl_{\weight}(\x)$, suggesting that the chi-squared bound with $\ell$ degrees of freedom is also appropriate in the $q$-dimensional subspace (depending on the rank of the variance matrix when emulator variance is included).
\par
When applying a basis rotation prior to calibration, we are ensuring that the representation of $\obs$ on the basis would not be ruled out (not in the terminal case). Theorem 1 further highlights the importance of this: if the reconstruction error given by basis $\bas_q$ is greater than the chi-squared bound, $\chi^2_{\ell, 0.995}$, then we are guaranteed to have an empty NROY space, as $\cimpl_{\weight}$ is non-negative. Checking whether a truncated basis passes this check is critical.

\subsection{Adding basis uncertainty}

When reconstructing a field from basis coefficient emulators, \citet{wilkinson2010bayesian} adds a variance term dependent on the discarded basis vectors to the posterior emulator variance. In this setting, the variance in \eqref{hmvar} becomes:
\begin{displaymath}
\text{Var}(\obs - \text{E}[f(\x)]) = \bas_q \text{Var}[\cc(\x)] \bas_q^T + \var_{\err} + \var_{\disc} + \bas_{-q} \boldsymbol{\Phi} \bas_{-q}^T,
\end{displaymath}
where $\bas_{-q}$ contains the remaining basis vectors from $\bas$, and $\boldsymbol{\Phi}$ is a diagonal matrix with entries corresponding to the eigenvalues of the columns of $\bas_{-q}$. This extra variance term is fixed, as are $\var_{\err}$ and $\var_{\disc}$, so that defining $\weight$ as:
\begin{displaymath}
\weight = \var_{\err} + \var_{\disc} + \bas_{-q} \boldsymbol{\Phi} \bas_{-q}^T,
\end{displaymath}
gives the same decomposition of $\fimpl$ as in Theorem 1.

\subsection{Relationship with univariate emulation} \label{sectionuv}

If we emulate grid boxes individually, we do not have the structure from Theorem 1, as it depends on having a basis $\bas_q$. Although in an univariate emulation approach, the $\ell \times \ell$ emulator variance $\text{Var}[f(\x)]$ is diagonal, $\var_{\err}$ and $\var_{\disc}$ will generally not be, hence there is an expensive inversion that varies with $\x$. We could instead use the univariate implausibility for each grid box individually, ignoring any correlations in the variance matrices, or match to global summaries of the output \citep{lee2016relationship,johnson2018importance}.
\par
\par
Another consequence of not having a low-dimensional basis is that it is not as straight-forward to assess whether the terminal case applies, i.e., whether $\obs$ is guaranteed to be ruled out. When using a basis, we can directly identify whether we are in the terminal case, by calculating the reconstruction error of $\obs$ on the basis, prior to emulation. This is not possible in the univariate case, as although we have $\ell$ degrees of freedom, we do not know whether the ensemble we have, and hence the emulators, will allow $\obs$ to be represented. 
\par
To discover whether we are in the terminal case with univariate emulation, we need to build $\ell$ emulators, and sample for a large Latin hypercube design across $\mathcal{X}$, assessing how close it is possible to get to $\obs$. The observed field may require extrapolation from the ensemble in several places, and it may not be possible to simultaneously achieve each of these extrapolations for some value of $\x$. This problem of not having enough ensemble signal to allow the directions of interest to be properly explored may also manifest in the basis emulation case, even if we are not in the terminal case (although this is likely better evidence that there is no $\best$ under the current error specification).
\par
A further drawback of independent univariate emulators is that because we have ignored any dependence across outputs, if we wish to draw a realisation of the field at $\x$ from the emulator posterior, the resulting field may not be smooth, as we may expect the true output to be. The basis approach will generally propagate smoothness into the posterior samples (see Figure \ref{emulatorplot} for a comparison of posterior samples for our climate application).

\section{Application to CanAM4} \label{sectionapplication}

In this section, we again consider the SLP output, and build emulators with three different methods:
\begin{enumerate}
\item Univariate emulators for each grid box (UV);
\item The SVD basis (SVD);
\item The optimally-rotated basis (ROT).
\end{enumerate}
We use the same proxy observations as in Section \ref{sectionprojection}, so that $\best$ is known, and $\var_{\disc} = 0$. For $\var_{\err}$, we use the form in \eqref{obserr}, with $\boldsymbol{\delta} = (5,5)$, $\sigma_i = 1/3$ outside of Antarctica (so that $\pm3\sigma = \pm1$), and $\sigma_i = 10/3$ in Antarctica.
\par
For consistency, when constructing emulators, whether for the univariate or a basis case, we use the RobustGaSP package \citep{gu2018robustgasp}. We initially fit emulators without a structured mean function, but allowing the correlation lengths to vary across the univariate emulators. Although this will take longer than estimating a common set of parameters, it gives the emulators greater flexibility. With truncation after 90\% of ensemble variability has been explained, the two basis methods require 12 (SVD) and 14 (ROT) vectors, and hence emulators.
\par
\begin{figure}[t]
\centering
\includegraphics[width = 0.8\linewidth]{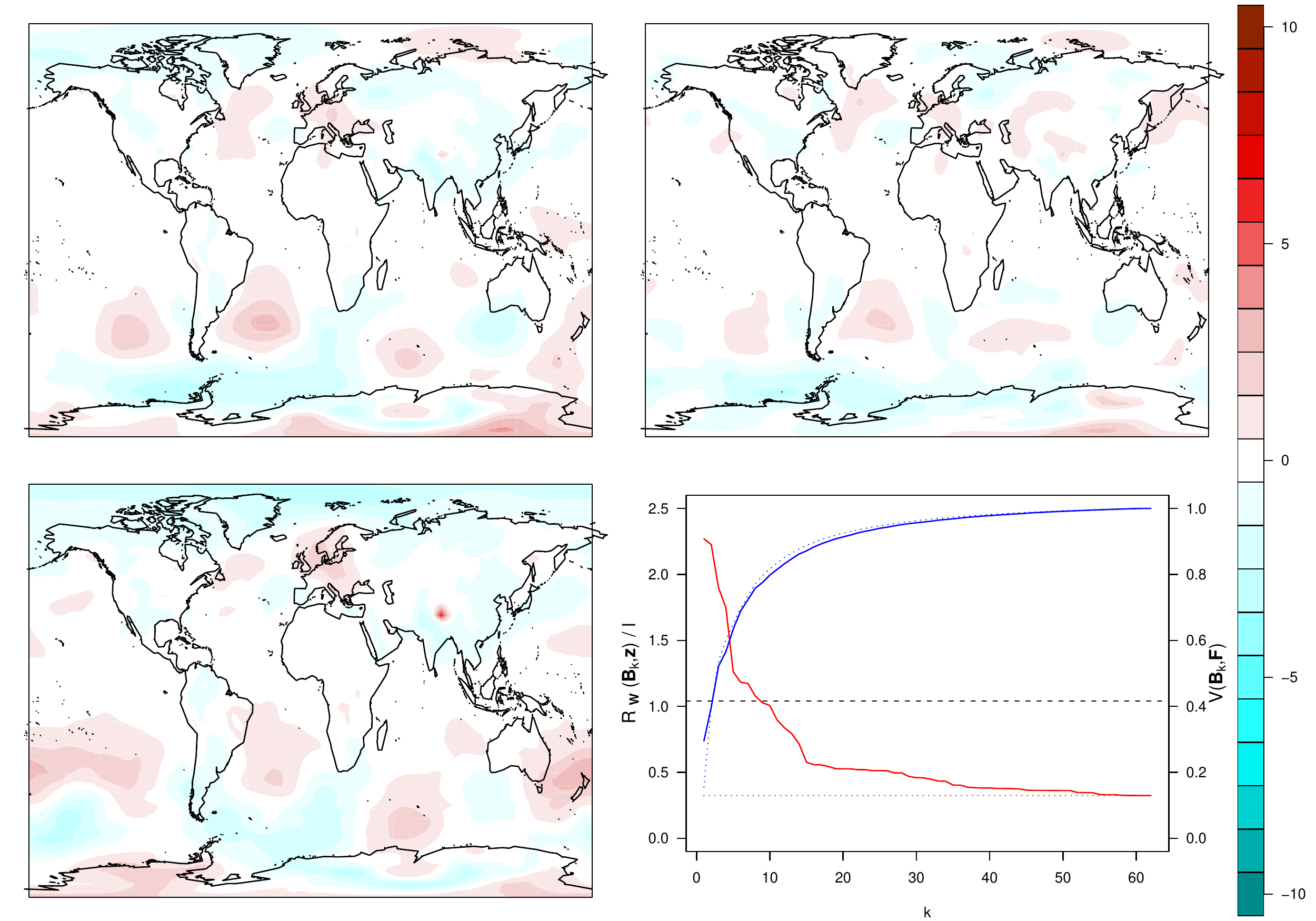}
\caption{The difference between $\obs$ and its reconstruction with the truncated SVD basis (top left), its reconstruction with the truncated ROT basis (top right), and its best match using the UV emulators (bottom left). The VarMSEplot in the final panel compares the SVD (solid lines) and ROT (dotted lines) bases, showing that the truncated ROT basis better represents $\obs$.}
\label{reconerrorplot}
\end{figure}

Prior to emulation, we assess how well each basis represents the observations, with the difference between $\obs$, and the truncated basis reconstructions, shown in the top half of Figure \ref{reconerrorplot}. From this, we see that the representation of $\obs$ is slightly more accurate for the rotated basis (always the case, as it explains as much of $\obs$ as possible given the ensemble), although here the SVD representation is generally close, with the majority of each plot coloured white. The VarMSEplot in the bottom right compares the reconstruction error (red lines) and variance explained (blue lines) for the truncated basis with $k$ vectors, for SVD (solid lines) and ROT (dotted lines). The ROT basis explains as much of $\obs$ as possible in the first basis vector here, and whilst the full SVD basis eventually represents $\obs$ equally well, when truncation occurs, the ROT basis is superior. Both truncated bases avoid the terminal case, with the reconstruction error below $\ibound$ (horizontal black dotted line).
\par
Theoretically, whilst the basis methods have a best possible representation, restricted by the choice of the basis, the univariate approach has full degrees of freedom, and can produce $\obs$ perfectly due to the independence of the emulators. In practice, this will not be the case, with extrapolation likely required in multiple locations to capture $\obs$ exactly. To assess whether the observations can be found with the univariate emulation method, we first construct emulators.

\subsection{Emulation}

We fit the 8192 UV, 12 SVD, and 14 ROT emulators, using the 62-member ensemble, with 49 runs from a separate ensemble reserved for out-of-sample validation (one of which is $f(\best)$).
\par
The closest representation of the observations given by the univariate emulators across $\x \in \mathcal{X}$ is shown in the third panel of Figure \ref{reconerrorplot}. This plot exhibits slightly larger, and more widespread, biases than the two basis versions, with the patterns generally similar to the SVD basis reconstruction. These are not directly comparable, as the basis plots assume perfect emulation is possible, and the problem of not having enough signal to accurately emulate around $\best$ may also be present for the basis methods, but this is the only way to assess whether we have the terminal case for the UV emulators (as discussed in Section \ref{sectionuv}).
\par
Using the 62 design and 49 validation runs, we compare the performances of each set of emulators, with Table \ref{emulatorval} showing several summary statistics for each. For consistency with projection and calibration, given that we have knowledge about how errors vary across the output, each summary is calculated with respect to the $\weight$ norm.
\par
For both in-sample (`Design') and out-of-sample (`Validation') performance, the two basis methods are more accurate than the UV emulation approach, with error reduced by 20\%-30\%. There is little difference between SVD and ROT here, although the prediction at $\best$ is more accurate when using the SVD emulators than ROT (with both more accurate than UV). Each gives a prediction at $\best$ that is not as close to $\obs$ as it is theoretically possible to achieve with the basis. As this is true for each method, this suggests it is a problem of extrapolation in the high-dimensional input space with a small design. Overall, there is little difference in the emulator accuracy for the two basis methods, with the UV method worse by these metrics.
\begin{table}[t]
\centering
\begin{tabular}{|c|c|c|c|}
\hline
\textbf{Type} & \textbf{UV} & \textbf{SVD} & \textbf{ROT} \\ \hline
Number of emulators & 8192 & 12 & 14 \\ \hline
Design & 1.544 & 1.194 & 1.153 \\ \hline
Validation & 2.362 & 1.657 & 1.644 \\ \hline
$R_{\weight}(\cdot, \obs) / \ell$ & 0 & 0.837 & 0.325 \\ \hline
$\lVert \obs - \text{E}[f(\best)] \rVert_{\weight} / \ell$ & 2.813 & 2.088 & 2.280 \\ \hline
$\mathcal{X}_{NROY}$ size & n/a & 47.31\% & 49.36\% \\ \hline
$\lVert \obs - \text{E}[f(\x)] \rVert_{\weight} / \ell, \,  \x \in \mathcal{X}_{NROY}$ & n/a & 2.531 & 2.417 \\ \hline
$\lVert \obs - \text{E}[f(\x)] \mathrm{exp}(-\cimpl_{\weight}(\x)) \rVert_{\weight} / \ell, \,  \x \in \mathcal{X}_{NROY}$ & n/a & 1.726 & 1.589 \\ \hline
\end{tabular}
\caption{Comparison of emulators. Each of the error statistics is in the $\weight$ norm, and scaled by $\ell$. The `Design' and `Validation' columns report the median error across the design and validation ensembles, $R_{\weight}(\cdot, \obs)$ gives how close it is possible to get to $\obs$, and $\lVert \obs - \text{E}[f(\best)] \rVert_{\weight}$ gives the emulator error at $\best$. The final two rows report the average difference between $\obs$ and emulator predictions for $\x \in \mathcal{X}_{NROY}$, both unweighted and weighted by $\exp(-\cimpl_{\weight}(\x))$.}
\label{emulatorval}
\end{table}

\par
We also fit linear (in the inputs) mean functions for the UV approach, which improved the emulator validation slightly (median error 2.017), but still performed worse than the constant mean basis emulators. The prediction at $\best$ was not improved by this new mean function. Fitting different mean functions for all $\ell = 8192$ emulators, prior to estimating the Gaussian process parameters, would give a further improvement, but we do not fit these here as the basis methods have proved to be more accurate, whilst also being significantly faster (see Section \ref{section_time}), for this application.
\par
When restricted to constant mean functions, SVD and ROT performed similarly (superior to UV). As we only need to fit a small number of emulators in each of these cases, we are able to spend time fitting new emulators with more complex mean functions. In this instance, adding structure to the mean did not give a significant improvement for either the SVD or ROT basis, and hence we proceed with the original emulators. However, needing to fit 12 or 14 emulators, instead of 8192, allows more time to be dedicated to each, and in general this will be more beneficial than an automated approach.
\par
\begin{figure}[t]
\centering
\includegraphics[width = 0.8\linewidth]{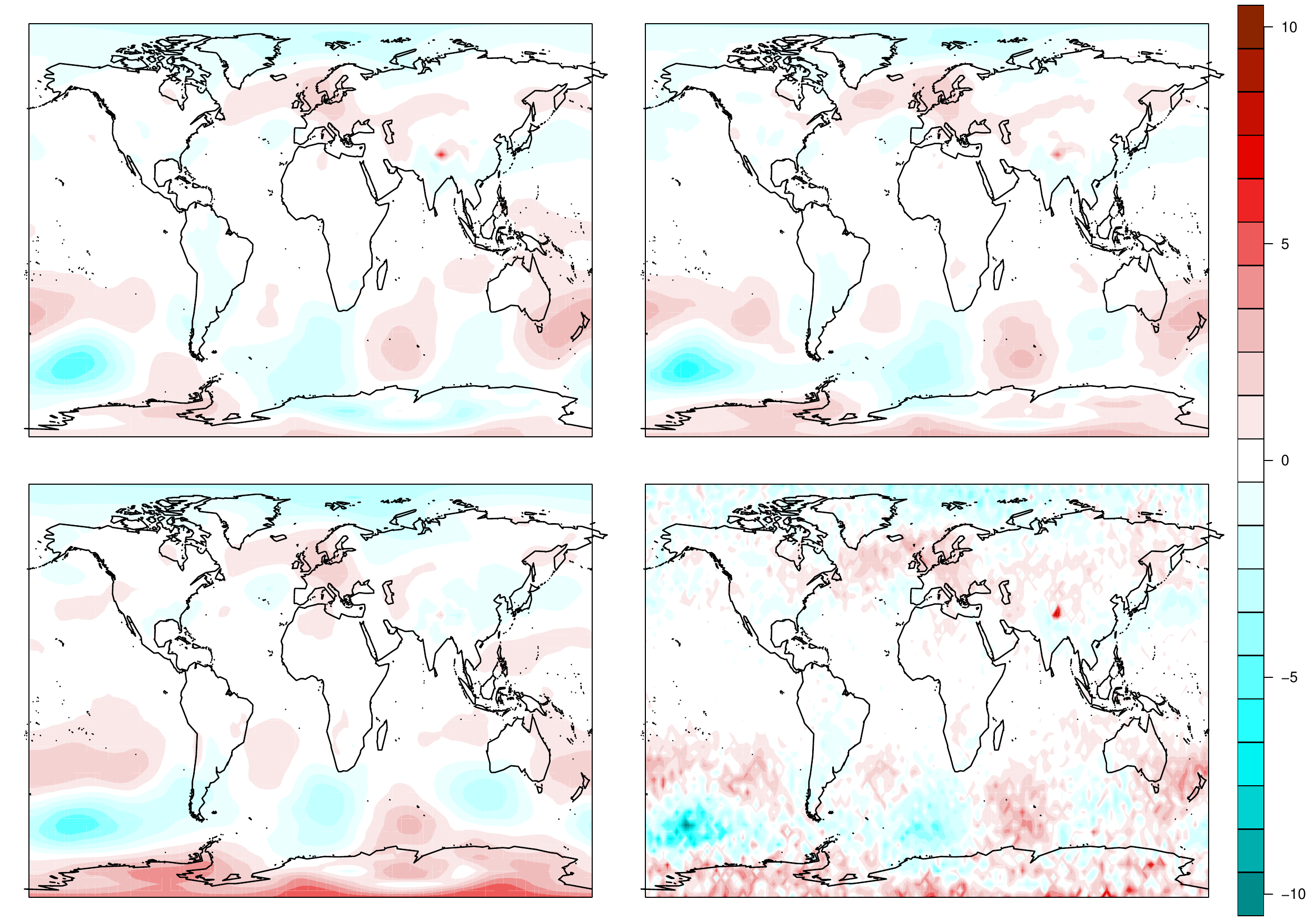}
\caption{Top: difference between $\obs$ and ROT (left), UV (right) emulator predictions at $\best$. Bottom: posterior samples from the ROT, UV emulators at $\best$.}
\label{emulatorplot}
\end{figure}

The top half of Figure \ref{emulatorplot} shows the difference between $\obs$ and the predicted fields at $\best$, for the ROT (left) and UV emulators (right). Visually, these two anomaly plots are reasonably similar, with biases in the same spatial locations. However, according to the $\weight$ norm, the prediction given by ROT is  closer to $\obs$ (2.280, compared to 2.813, from Table \ref{emulatorval}). The lower half of Figure \ref{emulatorplot} gives samples from the emulator posterior at $\best$ in each case. For both UV and ROT, the emulators are fitted independently, but samples from the ROT posterior (bottom left) retain smoothness, whereas for UV this is not the case.
\par
Each emulation method failing to get as close to the observations as theoretically possible shows that there may not be enough signal in the direction of $\obs$ in the 62 member ensemble to accurately emulate in this region of parameter space (or that other emulation methods beyond those considered here, e.g. non-stationary methods, may be required to improve accuracy). The inability to perfectly reproduce the observations suggests a wave of history matching, followed by a new design in NROY space, would be useful, if it were possible to run further ensembles of CanAM4.

\subsection{History matching}

As history matching large spatial fields is fast by exploiting the basis structure in Section \ref{sectionefficient}, given that we already have $\weight^{-1}$ and the reconstruction error, we now history match with the SVD and ROT emulators. We do not history match with the UV emulators, as they are less accurate here, and we do not have the required structure to enable fast calculation of $\fimpl$ (although we could instead match using a summary).
\par
\par
Each set of emulators fitted to the SLP data resulted in reasonably high variance, so that using $\ibound = \chi^2_{\ell, 0.995}$ rules out none of parameter space, hence for this application we set $\ibound = \chi^2_{q, 0.995}$, and rule out space with $\cimpl_{\weight}$ (as this is consistent with $\fimpl$, up to $R_{\weight}$). The resulting NROY spaces consist of 47.31\% (SVD) and 49.36\% (ROT) of $\mathcal{X}$.
\par
Although the two NROY spaces are similar in terms of size, they likely differ in their composition. The final two rows of Table \ref{emulatorval} give the difference between $\obs$ and emulator predictions for $\x \in \mathcal{X}_{NROY}$, with equal weightings across NROY space, and weighted by $\mathrm{exp}(-\cimpl_{\weight}(\x))$ (as a rough proxy for the likelihood in probabilistic calibration). In both cases, the fields in the ROT NROY space are more consistent with $\obs$.
\par
From Table \ref{emulatorval}, we see that the error at $\best$ is larger than the weighted error across NROY space, for both SVD and ROT. However, the emulator variance at $\best$ is substantially higher than average (due to being in a part of $\mathcal{X}$ unexplored by the design used to fit the emulators), so that whilst the mean predictive field is not as close as others in NROY space, the variance is higher, and $\best$ does have a relatively low implausibility.

\subsection{Computational time} \label{section_time}

\begin{table}[t]
\centering
\begin{tabular}{|c|c|c|c|c|c|}
\hline
Samples & SVD, $\weight^{-1}$ & $\text{E}[\cc(\x)]$ & $\fimpl(\x)$ & UV & ROT   \\ \hline
$10^3$ & 1492 & 0.13 & 6.01 & 74 & 1498 \\ \hline
$10^4$ & 1492 & 1.16 & 6.24 &  678 & 1499 \\ \hline
$10^5$ & 1492 & 11.46 & 8.71 & 6702 & 1512 \\ \hline
$10^6$ & 1492 & 106.88 & 30.94 & 62524 & 1630 \\ \hline
\end{tabular}
\caption{Time (in seconds) to evaluate quantities required for history matching, and a cost for the UV (emulator evaluation only) and ROT (SVD, $\weight^{-1}$, emulator evaluation, and implausibility calculation) methods.}
\label{timings_table}
\end{table}

We now quantify the savings afforded by the use of basis methods in this example. In order to build a picture of NROY space, we will generally need a minimum of 1 million samples from the emulator posterior. Fewer may be reasonable in some cases, e.g. if the input space has a lower dimension, but if this is not the first wave of history matching, many more may be required to identify a (potentially) small region of space that is not ruled out.
\par
The univariate approach requires 8192 emulator evaluations for each $\x$, compared to 14 for ROT, the more expensive of the two basis approaches in the previous section. To find $\text{E}[f(\x)]$ and $\text{Var}[f(\x)]$ at a sample of 1 million points, therefore, 14 million emulator evaluations are required for ROT, whilst 585 times more are needed for the UV approach. Given emulator expectations and variances, the implausibility for the basis method is also inexpensive to evaluate across a large sample, due to \eqref{implequation}. For UV, with the lack of structure, we do not have a fast method for $\fimpl$. 
\par
Table \ref{timings_table} compares the computational time required by the UV and ROT methods (using a MacBook Pro with 8GB memory, 2.3 GHz Intel Core i5 processor) when the number of samples from $\mathcal{X}$ increases. For the UV approach, we only include the cost of evaluating the 8192 emulators. For ROT, we include the fixed initial costs for the basis method (inverting $\weight$, calculating the basis), the emulator evaluations, and the implausibility calculations (including the one-off calculation of $\R_{\weight}(\bas_q, \obs)$).
\par
We see that the ROT method is significantly faster for $10^5$ or more samples, with little additional time required as the number of samples increases by a power of 10. A basis method has a larger initial cost, with the calculation of a) the SVD basis and b) the one-off inversion of $\weight$, but any subsequent matrix calculations (e.g. the reconstruction error) can exploit these stored quantities, and hence large savings are gained by the significantly fewer emulator evaluations required. Table \ref{timings_table} ignores the cost of the implausibility for the UV method, and even if we assume that rather than calculating the full implausibility, a fast summary is used to assess the resulting fields, and that this has no cost, the table shows that history matching for the ROT basis is significantly faster.
\par
Greater parallelisation, running on a faster machine, and potential simplifications may save more time for the UV method, however the scalability of the basis methods is unlikely to be surpassed. When history matching a climate model, we have many more fields, hence there is a greater benefit from having $O(10)$ emulators per field, rather than thousands, enabling more expert time to be spent on fitting each emulator, and exploration of $\mathcal{X}$ to proceed more efficiently. Calculating $\fimpl$ for each field is extremely fast (30 seconds for 1 million evaluations), whereas for UV emulators a summary would be needed to achieve this speed.

\section{Discussion} \label{discussion}

In this paper, we have shown that computer models with large output fields can be history matched efficiently, without any loss of information. The expensive implausibility is calculated for the entire $\ell$-dimensional field using only $q \times q$ ($q << \ell$) matrix inversions at each $\x$, exploiting consistency between the observation error and discrepancy variance matrices in the implausibility, and the weight matrix used for projection. We decomposed the implausibility over the field as the sum of the reconstruction error of the truncated basis, $\bas_q$, fixed for all $\x$, and a term dependent on the input parameters, the coefficient implausibility with projection in a certain norm.
\par
Projecting in $L_2$ (equal, uncorrelated weights on the $\ell$ outputs) or a structured weight can lead to different classification of points (history matching) and different distributions for $\best$ (probabilistic calibration), with more difference when there is more structure in $\weight$. However, as $\cimpl_{\weight}$ is perfectly correlated with $\fimpl$, projection in $\weight = \var_{\err} + \var_{\disc}$ is the appropriate choice, and we can calibrate over the original field, with no loss of information (assuming that the truncated basis is suitably representative of the full field).
\par
For the climate model example, both basis methods outperformed the univariate (UV) approach, according to all metrics considered. Theoretically, ROT can get closest to $\obs$, by its construction, although in our example, SVD and ROT performed similarly, whilst UV was unable to even get as close as SVD allows, which is likely to be the case generally when $\obs$ lies outside of the spread of the ensemble (often true for small ensembles, and particularly for climate examples).
\par
We showed that the basis methods offer far greater efficiency than the univariate approach in a calibration exercise, considering the fewer emulator evaluations needed, with savings for $>10^5$ samples from $\mathcal{X}$. Having emulation at least as accurate for the basis approaches as for the univariate case makes a basis method attractive, particularly when faced with multiple large output fields, as is commonly the case for climate models (finding the most physically-plausible field for a single output will likely lead to biases in others, so ideally all should be considered). The savings in having to emulate orders of magnitude fewer quantities (basis coefficients rather than grid boxes) allows more time to be spent constructing and rigorously validating emulators, with further savings given by the efficient implausibility calculations. Although we could history match to global summaries rather than the full field, this may hide competing biases.
\par
To apply a basis method, little extra work or knowledge is required, as standard univariate emulators can be fitted to basis coefficients, as in our application. Selecting an appropriate basis is therefore the main problem, and in many cases, an out-of-the-box method such as SVD, with a rotation when required, is fast and easy to apply, giving an intuitive spatial basis. Even if a summary is used for calibration, rather than the full field, so that the fast implausibility calculation demonstrated here is not required, building and evaluating fewer emulators gives computational savings, whilst yielding spatially-coherent samples from the emulator posterior, giving reason for utilising a basis approach.

\bibliographystyle{apa}
\bibliography{fullbib.bib}

\appendix

\section{Proof of Theorem 1}

We apply the Woodbury formula \citep{woodbury1950inverting, higham2002accuracy}:
\begin{equation}
(\textbf{A} + \textbf{UCV})^{-1} = \textbf{A}^{-1} - \textbf{A}^{-1} \textbf{U} (\textbf{C}^{-1} + \textbf{V} \textbf{A}^{-1} \textbf{U})^{-1} \textbf{V} \textbf{A}^{-1},
\end{equation}
where $\textbf{A}$ is an $\ell \times \ell$ matrix, $\textbf{C}$ is a $q \times q$ matrix, $\textbf{U}$ is an $\ell \times q$ matrix, and $\textbf{V}$ is a $q \times \ell$ matrix. 
\par
To prove the result, we show that the difference between the field implausibility and the reconstruction error can be written as $\cimpl_{\weight}$. We first expand the field implausibility using the Woodbury formula, so that:
\begin{align*}
\begin{split}
\fimpl(\x) &= (\obs - \bas_q \text{E}[\cc(\x)])^{T}(\bas_q \text{Var}[\cc(\x)] \bas_q^T + \weight)^{-1} (\obs - \bas_q \text{E}[\cc(\x)]) \\
&= (\obs - \bas_q \text{E}[\cc(\x)])^{T} \{ \weight^{-1} - \weight^{-1} \bas_q (\text{Var}[\cc(\x)]^{-1} + \bas_q^T \weight^{-1} \bas_q)^{-1} \bas_q^T \weight^{-1} \}
(\obs - \bas_q \text{E}[\cc(\x)]) \\
&= (\obs - \bas_q \text{E}[\cc(\x)])^{T} \weight^{-1} (\obs - \bas_q \text{E}[\cc(\x)]) \\
&- (\obs - \bas_q \text{E}[\cc(\x)])^{T} (\weight^{-1} \bas_q (\text{Var}[\cc(\x)]^{-1} + \mat)^{-1} \bas_q^T \weight^{-1}) (\obs - \bas_q \text{E}[\cc(\x)]), \\
\end{split}
\end{align*}
where $\mat = \bas_q^{T}\weight^{-1}\bas_q$. Applying the Woodbury formula again, we have:
\begin{displaymath}
(\text{Var}[\cc(\x)]^{-1} + \mat)^{-1} = \mat^{-1} - \mat^{-1} (\text{Var}[\cc(\x)] + \mat^{-1})^{-1} \mat^{-1}.
\end{displaymath}
Therefore, the field implausibility can be written as:
\begin{align} \label{fieldexpansion}
\begin{split}
\fimpl(\x) &= (\obs - \bas_q \text{E}[\cc(\x)])^{T} \weight^{-1} (\obs - \bas_q \text{E}[\cc(\x)]) \\
&- (\obs - \bas_q \text{E}[\cc(\x)])^{T} \weight^{-1} \bas_q \mat^{-1} \bas_q^T \weight^{-1} (\obs - \bas_q \text{E}[\cc(\x)]) \\
&+ (\obs - \bas_q \text{E}[\cc(\x)])^{T} \weight^{-1} \bas_q \mat^{-1} (\text{Var}[\cc(\x)] + \mat^{-1})^{-1} \mat^{-1} \bas_q^T \weight^{-1} (\obs - \bas_q \text{E}[\cc(\x)]). \\
\end{split}
\end{align}
By rewriting $\cimpl_{\weight}$ (from \eqref{coeffimpl}),
\begin{align} \label{coeffexpansion}
\begin{split}
\cimpl_{\weight}(\x) &= (\cc(\obs) - \text{E}[\cc(\x)])^{T}(\text{Var}[\cc(\x)] + \text{Var}[\cc(\err)] + \text{Var}[\cc(\disc)])^{-1} (\cc(\obs) - \text{E}[\cc(\x)]) \\
&= (\mat^{-1} \bas_q^T \weight^{-1} \obs - \text{E}[\cc(\x)])^{T}(\text{Var}[\cc(\x)] + \mat^{-1}\bas_q^T \weight^{-1} (\var_{\err} + \var_{\disc}) \weight^{-1} \bas_q\mat^{-1} )^{-1} \times \\
& \quad \, \, (\mat^{-1} \bas_q^T \weight^{-1} \obs - \text{E}[\cc(\x)]) \\
&= (\mat^{-1} \bas_q^T \weight^{-1} \obs - \text{E}[\cc(\x)])^{T}(\text{Var}[\cc(\x)] + \mat^{-1})^{-1} (\mat^{-1} \bas_q^T \weight^{-1} \obs - \text{E}[\cc(\x)]), \\
\end{split}
\end{align}
we have that the final line of \eqref{fieldexpansion} is the coefficient implausibility:
\begin{align*}
\begin{split}
&(\obs - \bas_q \text{E}[\cc(\x)])^{T} \weight^{-1} \bas_q \mat^{-1} (\text{Var}[\cc(\x)] + \mat^{-1})^{-1} \mat^{-1} \bas_q^T \weight^{-1} (\obs - \bas_q \text{E}[\cc(\x)]) \\
&= (\mat^{-1} \bas_q^T \weight^{-1} \obs - \mat^{-1} \bas_q^T \weight^{-1} \bas_q \text{E}[\cc(\x)])^{T} (\text{Var}[\cc(\x)] + \mat^{-1})^{-1} \times \\
& (\mat^{-1} \bas_q^T \weight^{-1} \obs - \mat^{-1} \bas_q^T \weight^{-1} \bas_q \text{E}[\cc(\x)])  \\
&= (\mat^{-1} \bas_q^T \weight^{-1} \obs - \text{E}[\cc(\x)])^{T}(\text{Var}[\cc(\x)] + \mat^{-1})^{-1} (\mat^{-1} \bas_q^T \weight^{-1} \obs - \text{E}[\cc(\x)]).
\end{split}
\end{align*}
Hence, from \eqref{fieldexpansion}, we have:
\begin{align} \label{fieldexpansion2}
\begin{split}
\fimpl(\x) &= (\obs - \bas_q \text{E}[\cc(\x)])^{T} \weight^{-1} (\obs - \bas_q \text{E}[\cc(\x)]) \\
&- (\obs - \bas_q \text{E}[\cc(\x)])^{T} \weight^{-1} \bas_q \mat^{-1} \bas_q^T \weight^{-1} (\obs - \bas_q \text{E}[\cc(\x)]) + \cimpl_{\weight}(\x). \\
\end{split}
\end{align}
Next, we rewrite the reconstruction error by adding and subtracting $\bas_q \text{E}[\cc(\x)]$:
\begin{align} \label{recondecomp}
\begin{split}
\R_{\weight}(\bas_q, \obs) &= (\obs - \bas_q \mat^{-1} \bas_q^{T} \weight^{-1} \obs)^T \weight^{-1} (\obs - \bas_q \mat^{-1} \bas_q^{T} \weight^{-1} \obs) \\
&= (\obs - \bas_q \text{E}[\cc(\x)] + \bas_q \text{E}[\cc(\x)] - \bas_q \mat^{-1} \bas_q^{T} \weight^{-1} \obs)^T \weight^{-1} \times \\
& (\obs - \bas_q \text{E}[\cc(\x)] + \bas_q \text{E}[\cc(\x)] - \bas_q \mat^{-1} \bas_q^{T} \weight^{-1} \obs) \\
&= (\obs - \bas_q \text{E}[\cc(\x)])^T \weight^{-1} (\obs - \bas_q \text{E}[\cc(\x)]) + \\
& (\bas_q \text{E}[\cc(\x)] - \bas_q \mat^{-1} \bas_q^{T} \weight^{-1} \obs)^T \weight^{-1} (\bas_q \text{E}[\cc(\x)] - \bas_q \mat^{-1} \bas_q^{T} \weight^{-1} \obs) + \\
& 2 (\obs - \bas_q \text{E}[\cc(\x)])^T \weight^{-1} (\bas_q \text{E}[\cc(\x)] - \bas_q \mat^{-1} \bas_q^{T} \weight^{-1} \obs) \\
&= \R_1 + \R_2 + \R_3.
\end{split}
\end{align}
$\R_1$ is already present in the decomposition of $\fimpl(\x)$ in \eqref{fieldexpansion2}. Using that:
\begin{displaymath}
\boldsymbol{\mathbb{I}} = \mat^{-1}\mat = \mat^{-1}\bas_q^T\weight^{-1}\bas_q,
\end{displaymath} 
we have:
\begin{align*}
\begin{split}
\R_2 &= (\bas_q \text{E}[\cc(\x)] - \bas_q \mat^{-1} \bas_q^{T} \weight^{-1} \obs)^T \weight^{-1} (\bas_q \text{E}[\cc(\x)] - \bas_q \mat^{-1} \bas_q^{T} \weight^{-1} \obs) \\
&= (\text{E}[\cc(\x)] - \mat^{-1} \bas_q^{T} \weight^{-1} \obs)^T \bas_q^T \weight^{-1} \bas_q (\text{E}[\cc(\x)] - \mat^{-1} \bas_q^{T} \weight^{-1} \obs) \\
&= (\mat^{-1}\bas_q^T\weight^{-1}\bas_q \text{E}[\cc(\x)] - \mat^{-1} \bas_q^{T} \weight^{-1} \obs)^T \mat (\mat^{-1}\bas_q^T\weight^{-1}\bas_q \text{E}[\cc(\x)] - \mat^{-1} \bas_q^{T} \weight^{-1} \obs) \\
&= (\bas_q \text{E}[\cc(\x)] - \obs)^T \weight^{-1} \bas_q \mat^{-1} \mat \mat^{-1}\bas_q^T\weight^{-1} (\bas_q \text{E}[\cc(\x)] - \obs) \\
&= (\bas_q \text{E}[\cc(\x)] - \obs)^T \weight^{-1} \bas_q \mat^{-1}\bas_q^T\weight^{-1} (\bas_q \text{E}[\cc(\x)] - \obs) \\
&= (\obs - \bas_q \text{E}[\cc(\x)])^T \weight^{-1} \bas_q \mat^{-1}\bas_q^T\weight^{-1} (\obs - \bas_q \text{E}[\cc(\x)]).
\end{split}
\end{align*}
Similarly,
\begin{align*}
\begin{split}
\R_3 &= 2 (\obs - \bas_q \text{E}[\cc(\x)])^T \weight^{-1} (\bas_q \text{E}[\cc(\x)] - \bas_q \mat^{-1} \bas_q^{T} \weight^{-1} \obs) \\
&= -2 (\obs - \bas_q \text{E}[\cc(\x)])^T \weight^{-1} \bas_q (\mat^{-1} \bas_q^{T} \weight^{-1} \obs - \text{E}[\cc(\x)]) \\
&= -2 (\obs - \bas_q \text{E}[\cc(\x)])^T \weight^{-1} \bas_q (\mat^{-1} \bas_q^{T} \weight^{-1} \obs - \mat^{-1}\bas_q^T\weight^{-1}\bas_q \text{E}[\cc(\x)]) \\
&= -2 (\obs - \bas_q \text{E}[\cc(\x)])^T \weight^{-1} \bas_q \mat^{-1} \bas_q^{T} \weight^{-1} (\obs - \bas_q \text{E}[\cc(\x)]). \\
\end{split}
\end{align*}
Hence, from \eqref{recondecomp}:
\begin{align*}
\begin{split}
\R_{\weight}(\bas_q, \obs) &= (\obs - \bas_q \text{E}[\cc(\x)])^T \weight^{-1} (\obs - \bas_q \text{E}[\cc(\x)])\\
&- (\obs - \bas_q \text{E}[\cc(\x)])^T \weight^{-1} \bas_q \mat^{-1} \bas_q^{T} \weight^{-1} (\obs - \bas_q \text{E}[\cc(\x)]), \\
\end{split}
\end{align*}
and combining this with \eqref{fieldexpansion2}, 
\begin{displaymath}
\fimpl(\x) = \R_{\weight}(\bas_q, \obs) + \cimpl_{\weight}(\x).
\end{displaymath}

\end{document}